\documentclass[12pt]{article}

\usepackage{graphicx}
\begin{document}

\begin{center}
{\bf On generalized ModMax model of nonlinear electrodynamics} \\
\vspace{5mm} S. I. Kruglov
\footnote{E-mail: serguei.krouglov@utoronto.ca}
\underline{}
\vspace{3mm}

\textit{Department of Physics, University of Toronto, \\60 St. Georges St.,
Toronto, ON M5S 1A7, Canada\\
Department of Chemical and Physical Sciences, University of Toronto,\\
3359 Mississauga Road North, Mississauga, Ontario L5L 1C6, Canada} \\
\vspace{5mm}
\end{center}

\begin{abstract}
A new generalized ModMax model of nonlinear electrodynamics with four parameters is proposed.
The ModMax model and Born--Infeld-type electrodynamics are particular cases of the present model.
It is shown that a singularity of the electric field at the center of point-like charged particles is absent. We found corrections to Coulomb’s law at $r\rightarrow\infty$ and obtained the total electrostatic and magnetic energies of point-like charges. Free electric and magnetic charges and their densities are obtained.
\end{abstract}


Recently, a new model, named ModMax model, of nonlinear electrodynamics (NED) which is duality and conformal invariant, as well as Maxwell electrodynamics, was proposed in \cite{Bandos}. Some aspects of this model and its applications were studies in \cite{Kosyakov,Habib,Bandos1,Townsend,Flores,Kubiznak}. Earlier proposals of conformal invariant NED was in \cite{Cembranos,Cembranos1,Denisov,Denisova}.


The duality-invariant conformal electrodynamics was introduced in \cite{Bandos} and it is described by the Lagrangian density
\begin{equation}
L=-{\cal F}\cosh(\gamma)+\sqrt{{\cal F}^2+{\cal G}^2}\sinh(\gamma),
\label{1}
\end{equation}
where
\begin{equation}
{\cal F}=\frac{1}{4}F_{\mu\nu}F^{\mu\nu}=\frac{1}{2}\left(\textbf{B}^2-\textbf{E}^2\right),~~~~{\cal G}=\frac{1}{4}F_{\mu\nu}\tilde{F}^{\mu\nu}=\textbf{B}\cdot \textbf{E},
\label{2}
\end{equation}
are Lorentz invariants with $\textbf{B}$, $\textbf{E}$ being the magnetic induction and electric fields correspondingly, $F_{\mu\nu}=\partial_\mu A_\nu-\partial_\nu A_\mu$, $\tilde{F}^{\mu\nu}=(1/2)\epsilon^{\mu\nu\alpha\beta}F_{\alpha\beta}$ is the dual electromagnetic field, and $\gamma$ is the dimensionless parameter.
This model as well as Maxwell electrodynamics with the Lagrangian density ${\cal L}_M=-{\cal F}$ possess singularities in the centre of point-like charges. In addition, the electromagnetic energy of charges is infinite. To smooth singularities we propose the generalized ModMax model with the Lagrangian density
\begin{equation}
{\cal L}=\frac{1}{\beta}\left(1-\left(1-\frac{\beta L}{\sigma}-\frac{\beta \lambda{\cal G}^2}{2\sigma}\right)^\sigma\right),
\label{3}
\end{equation}
where $L$ is given by Eq. (1), $\beta$ and $\lambda$ have the dimensions of $(length)^4$ and $\sigma$ is the dimensionless parameter. At $\sigma=1$, $\lambda=0$ we come to ModMax model, when $\gamma=0$ one has the Born--Infeld-type model \cite{Kruglov}, \cite{Kruglov1}, at $\sigma=1/2$, $\gamma=0$ we arrive at the generalized Born--Infeld model \cite{Kruglov2}, and at $\sigma=1/2$, $\gamma=0$, $\lambda=\beta$ one comes to Born--Infeld model ($\beta=1/b^2$) \cite{Born}. At $\sigma\rightarrow\infty$ the Lagrangian density (3) becomes
\begin{equation}
{\cal L}_{exp}=\frac{1}{\beta}\left(1-\exp\left(-\beta L-\beta \lambda{\cal G}^2/2\right)\right).
\label{4}
\end{equation}
Some exponential NED models were considered in \cite{Hendi}, \cite{Kruglov3}. Thus, our model (3) allows us to consider different NED by fixing the parameters introduced. The Born--Infeld model is of interest because at the low energy D-brain dynamics is governed by Born--Infeld-type
action \cite{Fradkin}, \cite{Tseytlin}. Making use of the Taylor series, at $\beta L\ll 1$, $\beta\lambda{\cal G}^2\ll 1$, the Lagrangian density (3) becomes
\begin{equation}
{\cal L}=L+\frac{\beta(1-\sigma)L^2}{2\sigma} + \frac{\lambda}{2}{\cal G}^2+{\cal O}(LG^2)+
{\cal O}(L^3).
\label{5}
\end{equation}
As a result, in the weak-field limit and small $\gamma$ when the condition $\beta L\ll 1$ is satisfied, the Lagrangian density (3) approaches to the ModMax model. At $\gamma=0$ Eq. (5) corresponds to the Heisenberg--Euler-type electrodynamics \cite{Kruglov4}.

Adding to Eq. (3) the source term $A_\mu j^\mu$ and varying the action we obtain the Euler--Lagrange equations
\begin{equation}
\partial_\mu\left({\cal L}_{\cal F}F^{\mu\nu}+{\cal L}_{\cal G}\tilde{F}^{\mu\nu}\right)=j^\nu,
\label{6}
\end{equation}
where
\[
{\cal L}_{\cal F}=\frac{\partial {\cal L}}{\partial {\cal F}}=\left(1-\frac{\beta L}{\sigma}-\frac{\beta \lambda{\cal G}^2}{2\sigma}\right)^{\sigma-1}L_{\cal F},~~~
L_{\cal F}=-\cosh(\gamma)+\frac{{\cal F}\sinh(\gamma)}{\sqrt{{\cal F}^2+{\cal G}^2}},
\]
\begin{equation}
{\cal L}_{\cal G}=\frac{\partial {\cal L}}{\partial {\cal G}}=\left(1-\frac{\beta L}{\sigma}-\frac{\beta \lambda{\cal G}^2}{2\sigma}\right)^{\sigma-1}(L_{\cal G}+\lambda G),~~~
L_{\cal G}=\frac{{\cal G}\sinh(\gamma)}{\sqrt{{\cal F}^2+{\cal G}^2}}.
\label{7}
\end{equation}
Field equations (6) can be represented as Maxwell equations
making use of definitions of the electric displacement and magnetic fields
\[
\textbf{D}=\frac{\partial {\cal L}}{\partial \textbf{E}}=\varepsilon \textbf{E}+\nu\textbf{B},~~~\varepsilon=-{\cal L}_{\cal F},~\nu={\cal L}_{\cal G},
\]
\begin{equation}
\textbf{H}=-\frac{\partial {\cal L}}{\partial \textbf{B}}=\mu^{-1}\textbf{B}-\nu \textbf{E},~~~\mu=\varepsilon^{-1}.
\label{8}
\end{equation}
With help of Eq. (8) Euler--Lagrange equations (6) can be represented as Maxwell equations in Gaussian quantities
\begin{equation}
\nabla\cdot\textbf{D}=4\pi\rho,~~\frac{\partial\textbf{D}}{\partial t}-\nabla\times \textbf{H}=4\pi\textbf{j}.
\label{9}
\end{equation}
Second pair of Maxwell equations follows from the Bianchi identity $\partial_\mu\tilde{F}^{\mu\nu}=0$,
\begin{equation}
\nabla\cdot\textbf{B}=0,~~\frac{\partial\textbf{B}}{\partial t}+\nabla\times\textbf{E}=0.
\label{10}
\end{equation}
From Eq. (8) we obtain the relation
\begin{equation}
\textbf{D}\cdot\textbf{H}=(\varepsilon^2-\nu^2){\cal G}+2\varepsilon\nu {\cal F}.
\label{11}
\end{equation}
According to the criterion of \cite{Gibbons} the dual symmetry takes place if $\textbf{D}\cdot\textbf{H}=\textbf{E}\cdot\textbf{B}$. One can verify from Eq. (11) that the dual symmetry holds in two cases: $\sigma=1$, $\lambda=0$ which corresponds to ModMax model or for Born--Infeld-type model with $\sigma=1/2$, $\lambda=\beta$. It is worth noting that the two-parametric generalized Born--Infeld model ($\sigma=1/2$, $\lambda=\beta$) was considered and shown to be duality invariant in \cite{Bandos1}.


From Eq. (9), for the source of the point-like charged particle with the electric charge $q$, in Gaussian units, we obtain the equation as follows
\begin{equation}
\nabla\cdot\textbf{D}=4\pi q\delta(\textbf{r}),
\label{12}
\end{equation}
with the solution
\begin{equation}
\textbf{D}=\frac{ q\textbf{r}}{r^3}.
\label{13}
\end{equation}
From Eqs. (8) and (13) one finds
\begin{equation}
E\exp(\gamma)\left(1-\frac{\beta E^2}{2\sigma}\exp(\gamma)\right)^{\sigma-1}=\frac{ q}{r^2}.
\label{14}
\end{equation}
Introducing the dimensionless variables
\begin{equation}
v=\frac{\sqrt{\beta} E}{\sqrt{2\sigma}}\exp\left(\frac{\gamma}{2}\right),~~~u=\frac{r\sqrt[4]{2\sigma}}{{\sqrt[4]\beta} \sqrt{q}}\exp\left(\frac{\gamma}{4}\right),
\label{15}
\end{equation}
Eq. (14) takes the form
\begin{equation}
v\left(1-v^2\right)^{\sigma-1}=\frac{1}{u^2}.
\label{16}
\end{equation}
It follows from Eq. (16) that at $u\rightarrow 0$, $v(0)=1$ for $\sigma<1$, or in terms of electric fields
\begin{equation}
E(0)=\sqrt{\frac{2\sigma}{\beta}}\exp\left(-\frac{\gamma}{2}\right).
\label{17}
\end{equation}
Thus, the electric field of the point-like charged particle in the center is finite and possesses the maximum value for $\sigma<1$. When $\sigma$ increases the maximum value of electric fields increases, but if $\gamma$  increases the electric field in the center decreases.
For the cases $\sigma=1/2$ (Born--Infeld-type model) and $\sigma=3/4$ exact solutions to Eq. (16) and their asymptotic as $u\rightarrow 0$ are
\[
v=\frac{1}{\sqrt{1+u^4}},~~~~v=1-\frac{u^4}{2}+\frac{3u^8}{8}+{\cal O}(u^{12})~~~~\sigma=\frac{1}{2},
\]
\begin{equation}
v=\frac{\sqrt{\sqrt{1+4u^8}-1}}{\sqrt{2}u^4},~~~~v=1-\frac{u^8}{2}  +{\cal O}(u^{16})~~~~\sigma=\frac{3}{4}.
\label{18}
\end{equation}
The plots of function $v(u)$ for $\sigma=1/4$, $\sigma=1/2$ and $\sigma=3/4$ are given in Fig. (1).
\begin{figure}[h]
\includegraphics[height=4.0in,width=4.0in]{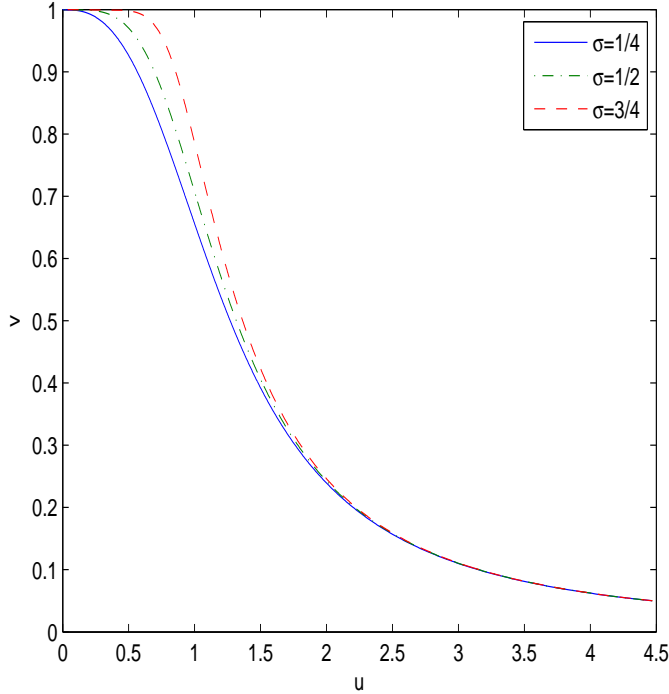}
\caption{\label{fig.1}The plot of the function $v(u)$ for $\sigma=1/4$, $\sigma=1/2$ and $\sigma=3/4$.}
\end{figure}
In the general case for $\sigma<1$, the functions $v(u)$ as $u\rightarrow 0$ ($r\rightarrow 0$) and $u\rightarrow \infty$ ($r\rightarrow\infty$) become
\[
v=1-\frac{u^{2/(1-\sigma)}}{2}+{\cal O}(u^{4/(1-\sigma)})~~~~u\rightarrow 0,
\]
\begin{equation}
v=\frac{1}{u^2}-\frac{1-\sigma}{u^6}+{\cal O}(u^{-8})~~~~u\rightarrow\infty.
\label{19}
\end{equation}
Making use of Eqs. (15) and (19) we obtain the asymptotic value of the electric field as $r\rightarrow 0$ and $r\rightarrow\infty$
\[
E=\sqrt{\frac{2\sigma}{\beta}}\exp\left(-\frac{\gamma}{2}\right)-\frac{(2\sigma)^{(2-\sigma)/(2(1-\sigma))}r^{2/(1-\sigma)}}
{2\beta^{(2-\sigma)/(2(1-\sigma))}q^{1/(1-\sigma)}}\exp\left(\frac{\gamma\sigma}{(2(1-\sigma)}\right)
\]
\[
+{\cal O}(r^{4/(1-\sigma))})~~~~~~r\rightarrow 0,
\]
\begin{equation}
E=\frac{q}{r^2}\exp(-\gamma)-\frac{(1-\sigma)\beta q^3}{2\sigma r^6}\exp(-2\gamma)+{\cal O}(r^{-8})~~~~~~r\rightarrow\infty.
\label{20}
\end{equation}
Equation (20) gives the correction to Coulomb's law as $r\rightarrow\infty$. We have damping of the electric field because of parameter $\gamma$.
Corrections to Coulomb’s law for the case of Born--Infeld-type electrodynamics ($\sigma=1/2$) and exponential-like electrodynamics (4) ($\sigma=\infty$) are similar but with the opposite sign. At $\sigma=1$, $\lambda=0$, $\gamma=0$, one has Maxwell’s electrodynamics and we come to the Coulomb law $E=q/r^2$ as $r\rightarrow\infty$, but the electric field at the origin is infinite.


The energy-momentum tensor is given by

\begin{equation}
T_{\mu\nu}=F_{\mu\alpha}\left({\cal L}_{\cal F}F_{~\nu}^{\alpha}+{\cal L}_{\cal G}\tilde{F}_{~\nu}^{\alpha}\right)-g_{\mu\nu}{\cal L}.
\label{21}
\end{equation}
From Eq. (21) we obtain the energy density
\begin{equation}
T_{0}^{~0}=-E^2{\cal L}_{\cal F}-G{\cal L}_{\cal G}-{\cal L}.
\label{22}
\end{equation}
In the case of pure electric field ($\textbf{B}=0$) the energy density (22) becomes
\begin{equation}
T_{0}^{~0}=E^2e^\gamma\left(1-\frac{\beta E^2}{2\sigma}e^\gamma\right)^{\sigma-1}-\frac{1}{\beta}\left(1-\left(1-\frac{\beta E^2}{2\sigma}e^\gamma\right)^\sigma\right).
\label{23}
\end{equation}
For the Born--Infeld-type electrodynamics with $\sigma=1/2$ we can calculate the total electrostatics energy of point-like charged particles. Introducing the dimensionless variables
\begin{equation}
y=\beta E^2\exp(\gamma),~~~~x=\frac{r^2\exp(\gamma/2)}{q\sqrt{\beta}},
\label{24}
\end{equation}
we obtain from Eq. (14) equation
\begin{equation}
y=\frac{1}{1+x^2}.
\label{25}
\end{equation}
Then, making use of Eqs. (23), (24) and (25), the total electrostatics energy of point-like charged particles (for $\sigma=1/2$) is given by
\[
{\cal E}=4\pi\int_0^\infty T_{0}^{~0}r^2dr=\frac{2\pi q^{3/2}\exp(-3\gamma/4)}{\beta^{1/4}}I,
\]
\begin{equation}
I=\int_0^\infty\left(\frac{\sqrt{1+x^2}}{\sqrt{x}}-\sqrt{x}\right)dx=-\frac{2}{3}\sqrt{x}\left(x-3 {_2}F_1\left(-\frac{1}{2},\frac{1}{4};\frac{5}{4};-x^2\right)\right)_0^\infty.
\label{26}
\end{equation}
Taking into account the asymptotic of hypergeometric function  $_2F_1(-1/2,1/4;5/4;-x^2)$ we obtain the total electrostatics energy of point-like charged particles
\begin{equation}
{\cal E}\approx 15.5327\frac{q^{3/2}}{\beta^{1/4}}\exp\left(-\frac{3\gamma}{4}\right).
\label{27}
\end{equation}
For Born--Infeld electrodynamics ($\beta=1/b^2$) at $\gamma=0$ we come to the result obtained in \cite{Born}.

To study the conformal invariance of the proposed model (3), we calculate the trace of the energy-momentum tensor $T_\mu^{~\mu}$. According to \cite{Hagen}, the conformal invariance takes place if $T_\mu^{~\mu}=0$. From Eqs. (7) and (27) we obtain
\[
T_\mu^{~\mu}=4\left({\cal F}{\cal L}_{\cal F}+{\cal G}{\cal L}_{\cal G}-{\cal L}\right)
\]
\begin{equation}
=4\left(1-\frac{\beta L}{\sigma}-\frac{\beta \lambda{\cal G}^2}{2\sigma}\right)^{\sigma-1}\left(\frac{(\sigma-1)L}{\sigma}+\frac{(2\sigma-1)\lambda G^2}{2\sigma}+\frac{1}{\beta}\right)-\frac{4}{\beta}.
\label{28}
\end{equation}
Equation (28) shows that the conformal invariance ($T_\mu^{~\mu}=0$) holds only for $\sigma=1$, $\lambda=0$ corresponding to the ModMax model.


In accordance with \cite{Born} we introduce the ``free charge density" $\rho_{free}$ by the equation
\begin{equation}
\nabla\cdot E=\frac{1}{r^2}\frac{d}{dr}(r^2E)=4\pi\rho_{free},
\label{29}
\end{equation}
where $E$ obeys Eq. (14). To calculate the distribution of the free charge one has to obtain $\rho_{free}$ from Eq. (29). We have exact solutions to Eq. (14) for $\sigma=1/2$ and $\sigma=3/4$. Making use of Eq. (15), from Eq. (18) we find
\[
E=\frac{q\exp(-\gamma/2)}{r_0^2\sqrt{1+(r/r_0)^4\exp(\gamma)}}~~~~~~~~\sigma=\frac{1}{2},
\]
\begin{equation}
E=\frac{qr_0^2\exp(-3\gamma/2)\sqrt{\sqrt{1+(9r^8/r_0^8)\exp(2\gamma)}-1}}{\sqrt{3}r^4}~~~~~~~~\sigma=\frac{3}{4},
\label{30}
\end{equation}
where $r_0=\sqrt[4]{\beta}\sqrt{q}$. When $q$ is the charge of the electron, $r_0$ is the electron radius \cite{Born}. It follows from Eq. (30) that for $r\rightarrow\infty$ the electric field $E\rightarrow (q/r^2)\exp(-\gamma)$ and for $r\rightarrow 0$ we have
\[
 E\rightarrow \frac{1}{\sqrt{\beta}}\exp\left(-\frac{\gamma}{2}\right)~~~~~~~~~\sigma=\frac{1}{2},
\]
 \begin{equation}
  E\rightarrow \sqrt{\frac{3}{2\beta}}\exp\left(-\frac{\gamma}{2}\right)~~~~~~~~~\sigma=\frac{3}{4},
 \label{31}
\end{equation}
 according to Eq. (20). From Eqs. (29) and (30) we obtain ``free charge density"
 \[
 \rho_{free}=\frac{q\exp(-\gamma/2)}{2\pi r_0^2r(1+(r/r_0)^4\exp(\gamma))^{3/2}}    ~~~~~~~~~\sigma=\frac{1}{2},
 \]
\begin{equation}
\rho_{free}=\frac{qr_0^2\exp(-3\gamma/2)}{2\sqrt{3}\pi r^5}\sqrt{\frac{\sqrt{1+(9r^8/r_0^8)\exp(2\gamma)}-1}{1+(9r^8/r_0^8)\exp(2\gamma)}}   ~~~~~~~~~\sigma=\frac{3}{4},
\label{32}
\end{equation}
For Born--Infeld electrodynamics, in the case $\sigma=1/2$, $\gamma=0$, one comes from Eq. (32) to the result found in \cite{Born}. Now we can calculate free charges
\begin{equation}
q_{free}=4\pi\int_0^\infty r^2\rho_{free}dr=\left(r^2E^2\right)_0^\infty.
\label{33}
\end{equation}
Making use of Eqs. (20) and (33) we find the free charge for any parameter $\sigma$
\begin{equation}
q_{free}=q\exp(-\gamma).
\label{34}
\end{equation}
Equation (34) shows that the free charge is less from $q$ by the factor $\exp(-\gamma)$.
From Eq. (30) we obtain the free electric charges, in the cases $\sigma=1/2$ and $\sigma=3/4$, inside the sphere $r<r_0$
\[
q_{free}(r_0)=4\pi\int_0^{r_0} r^2\rho_{free}dr=\left(r^2E\right)_0^{r_0}=q_{free}\frac{\exp(\gamma/2)}{\sqrt{1+\exp(\gamma)}}~~~~~\sigma=1/2,
\]
\begin{equation}
q_{free}(r_0)=q_{free}\frac{\sqrt{\sqrt{(1+9\exp(2\gamma)}-1}}{\sqrt{3}}\exp\left(-\frac{\gamma}{2}\right)  ~~~~~~~~~~~~~\sigma=3/4
\label{35}
\end{equation}
The plots of the functions $q_{free}(r_0)/q_{free}$ versus $\gamma$ is presented in Fig. 2.
\begin{figure}[h]
\includegraphics[height=4.0in,width=4.0in]{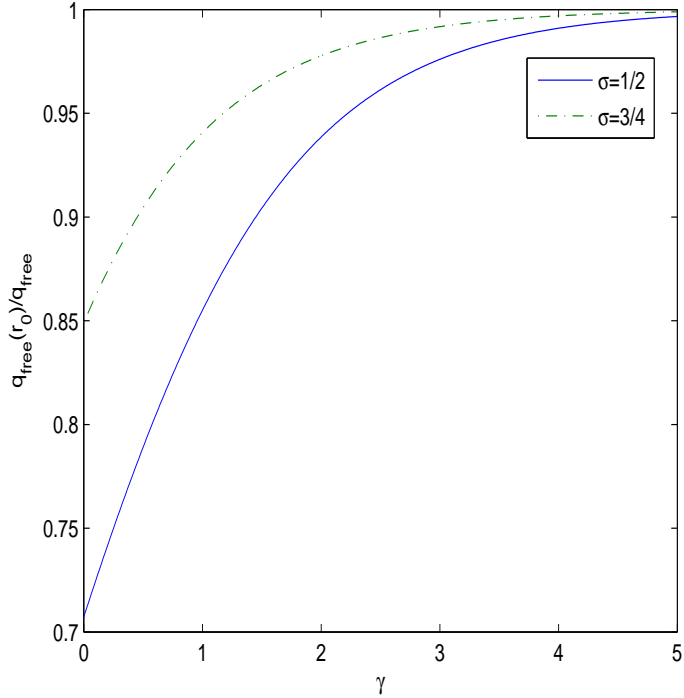}
\caption{\label{fig.2}The plot of the functions $q_{free}(r_0)/q_{free}$ versus $\gamma$ for $\sigma=1/2$ and $\sigma=3/4$.}
\end{figure}
It follows from Eq. (35) and the Fig. 2 that at $\gamma=0$ $q_{free}(r_0)/q_{free}\approx 0.71$ for $\sigma=1/2$ and $q_{free}(r_0)/q_{free}\approx 0.85$ for $\sigma=3/4$. Therefore, $71\%$ of the electron charge $q_{free}$ is contained in the electron radius sphere  for $\sigma=1/2$ and $85\%$ of the electron charge is concentrated within the electron radius for $\sigma=3/4$. When parameter $\gamma$ increases
more charge is inside the electron radius sphere.


For the magnetic monopole we have equation
\begin{equation}
\nabla\cdot\textbf{B}=4\pi Q\delta(\textbf{r}),
\label{36}
\end{equation}
where $Q$ is the magnetic charge. For a magnetic monopole we have equation
\begin{equation}
B=\frac{Q}{r^2}.
\label{37}
\end{equation}
From Eq. (22) we obtain the magnetic energy density ($E=0$) of the magnetic monopole
\begin{equation}
T_{0}^{~0}=-{\cal L}=\frac{1}{\beta}\left(\left(1+\frac{\beta Q^2}{2\sigma r^4}e^{-\gamma}\right)^\sigma-1\right).
\label{38}
\end{equation}
The total magnetic energy is given by
\begin{equation}
{\cal E}_m=4\pi\int_0^\infty T_{0}^{~0}r^2dr=-\frac{4\pi}{\beta}\int_0^\infty r^2\left(1-\left(1+\frac{\beta Q^2}{2\sigma r^4}e^{-\gamma}\right)^\sigma\right)dr
\label{39}
\end{equation}
Integral (39) converges for $0<\sigma<1$ and the energy of the magnetic monopole is finite within our model. In Table 1 we present the approximate values of the dimensionless energy $\bar{\cal E}_m=\beta^{1/4}Q^{-3/2}\exp(3\gamma/4){\cal E}_m$ for $\sigma=0.1,0.2,..,0.7$.
\begin{table}[ht]
\caption{The dimensionless energy $\bar{{\cal E}}_m$}
\centering
\begin{tabular}{c c c c c c c c}\\[1ex]
\hline
$\sigma$ & 0.1 & 0.2 & 0.3 & 0.4 & 0.5 & 0.6 & 0.7  \\[0.5ex]
\hline
 $\bar{\cal E}_m$ & 6.58 & 8.38 & 10.13 & 12.28 & 15.45 & 22.29 & 58.78 \\[0.5ex]
\hline
\end{tabular}
\end{table}
Table 1 shows that with increasing parameter $\sigma$ the magnetic energy of the monopole increasing. When parameter $\gamma$ increases the magnetic energy ${\cal E}_m$ decreases.

From Eq. (8) we obtain the magnetic field of the monopole
\begin{equation}
\textbf{H}=\frac{Q}{r^2}e^{-\gamma}\left(1+\frac{\beta Q^2}{2\sigma r^4}e^{-\gamma}\right)^{\sigma-1}.
\label{40}
\end{equation}
The ``free magnetic charge density" $\eta_{free}$ is defined by the equation
\begin{equation}
\nabla\cdot H=\frac{1}{r^2}\frac{d}{dr}(r^2H)=4\pi\eta_{free}.
\label{41}
\end{equation}
Making use of Eq. (41) one finds from Eq. (40) the ``free magnetic charge density"
\begin{equation}
\eta_{free}=-\frac{\beta Q^3(\sigma-1)}{2\pi \sigma r^7}e^{-2\gamma}\left(1+\frac{\beta Q^2}{2\sigma r^4}e^{-\gamma}\right)^{\sigma-2}.
\label{42}
\end{equation}
Similar to Eq. (33) we obtain, by using Eq. (40), the free magnetic charge
\begin{equation}
Q_{free}=\frac{1}{4\pi}\int \nabla\cdot Hd^3x=\int_0^\infty\frac{d}{dr}(r^2H)dr=\left(r^2H^2\right)_0^\infty=Q\exp(-\gamma).
\label{43}
\end{equation}
Thus, the equation for the free magnetic charge is similar to the free electric charge (34).
From Eq. (40), one finds the free magnetic charge inside the sphere of the radius $r_m=\beta^{1/4}\sqrt{Q}$
\begin{equation}
Q_{free}(r_m)=\left(r^2H\right)_0^{r_0}=Q_{free}\left(1+\frac{1}{2\sigma}e^{-\gamma}\right)^{\sigma-1}.
\label{44}
\end{equation}
The plots of the functions $Q_{free}(r_m)/Q_{free}$ versus $\gamma$ for $\sigma=1/8$, $\sigma=1/2$ and $\sigma=3/4$ is presented in Fig. 3.
\begin{figure}[h]
\includegraphics[height=4.0in,width=4.0in]{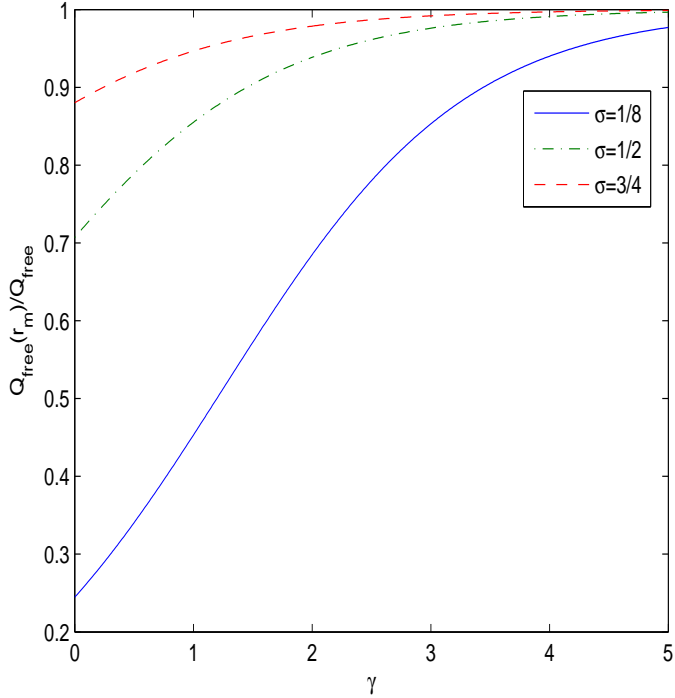}
\caption{\label{fig.3}The plot of the functions $Q_{free}(r_m)/Q_{free}$ versus $\gamma$ for $\sigma=1/8$, $\sigma=1/2$ and $\sigma=3/4$.}
\end{figure}
According to Eq. (44) $(1+e^{-\gamma}/(2\sigma))^{\sigma-1}$ of the magnetic charge $Q_{free}$ is inside the sphere of the radius $r_m$. The plot of the function $Q_{free}(r_m)/Q_{free}$ is in Fig. 4.
\begin{figure}[h]
\includegraphics[height=4.0in,width=4.0in]{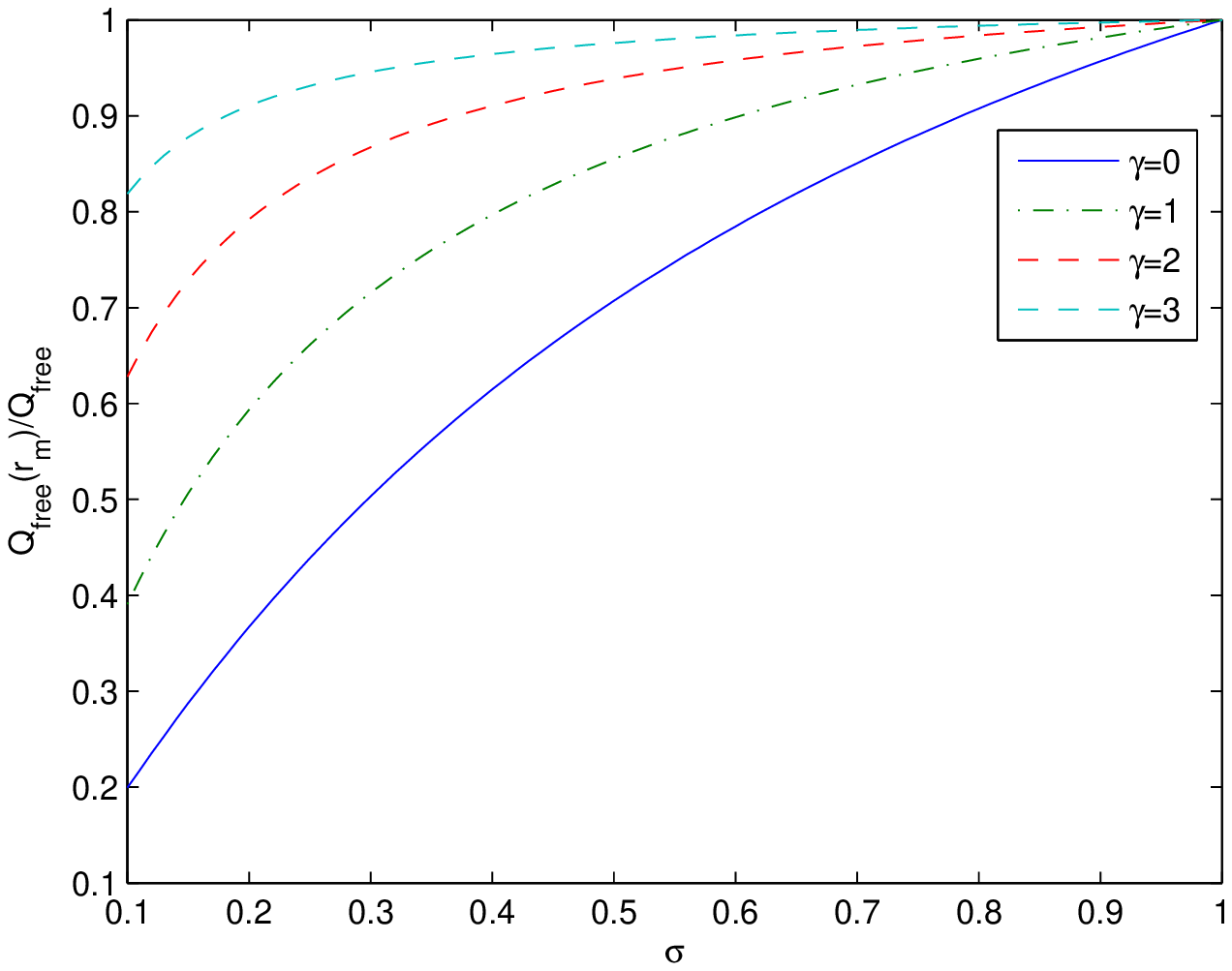}
\caption{\label{fig.4}The plot of the functions $Q_{free}(r_m)/Q_{free}$ versus $\sigma$, for $\gamma=0,1,2,3$.}
\end{figure}
Figure 4 shows that the ratio $Q_{free}(r_m)/Q_{free}$ increases with increasing parameters $\sigma$ and $\gamma$. As a result, more magnetic charge is concentrated in the sphere of the radius $r_m$ with greater values of $\sigma$ and $\gamma$.


The generalised ModMax model proposed as compared to ModMax model (1) possesses the attractive features as follows.

$\bullet$ At some parameters $\beta$, $\lambda$, $\gamma$ and $\sigma$ we come to different models (including the Born--Infeld electrodynamics) discussed in the literature.

$\bullet$ In the weak-field limit and $\gamma=0$, the Lagrangian density leads to the Heisenberg--Euler electrodynamics.

$\bullet$ The electric field of the point-like charged particle in the origin is
finite and possesses the maximum value (for $\sigma<1$).

$\bullet$ The electric and magnetic energies of point-like charged particles is finite for some parameters $\sigma$.

Such properties take place also for the two-parametric duality invariant
generalization of  Born--Infeld electrodynamics found in \cite{Bandos}, \cite{Bandos1}.

In addition, we calculated the free electric and magnetic densities in our model that allow us to study the distributes of the electric and magnetic charges in the space.

\end{document}